\documentclass[aps,prb,twocolumn,amsmath,amssymb,superscriptaddress,citeautoscript]{revtex4-2}
\usepackage{graphicx}% Include figure files
\usepackage{dcolumn}% Align table columns on decimal point
\usepackage{bm}% bold math
\usepackage{latexsym,epsfig}
\usepackage{graphicx}
\usepackage[export]{adjustbox}
\usepackage{verbatim}
\usepackage{comment}
\usepackage{amsmath}
\usepackage{amssymb}
\usepackage{stmaryrd}
\usepackage{color}
\usepackage{epstopdf}
\usepackage{grffile}
\usepackage{mathtools}
\usepackage{physics}
\usepackage{ulem}
\DeclareGraphicsExtensions{.eps}

\usepackage{hyperref}
\hypersetup{
    colorlinks=true,      
    urlcolor=blue,
    citecolor=blue,
    linkcolor=blue
}

\begin{document}
\title{Enhanced many-body localization in a kinetically  constrained model}
\author{Karl Royen} 
\affiliation{Institut f\"{u}r Theoretische Physik, Georg-August-Universit\"{a}t G\"{o}ttingen, D-37077 G\"{o}ttingen, Germany}
\author{Suman Mondal} 
\affiliation{Institut f\"{u}r Theoretische Physik, Georg-August-Universit\"{a}t G\"{o}ttingen, D-37077 G\"{o}ttingen, Germany}
\author{Frank Pollmann}
\affiliation{Technical University of Munich, TUM School of Natural Sciences, Physics Department, 85748 Garching, Germany} 
\affiliation{Munich Center for Quantum Science and Technology (MCQST), Schellingstr. 4, 80799 M\"unchen, Germany}
\author{Fabian Heidrich-Meisner}
\affiliation{Institut f\"{u}r Theoretische Physik, Georg-August-Universit\"{a}t G\"{o}ttingen, D-37077 G\"{o}ttingen, Germany}

\date{\today}

\begin{abstract}
In the study of the thermalization of closed quantum systems, the 
role of kinetic constraints on the temporal dynamics and the eventual 
thermalization is attracting significant interest. Kinetic constraints
typically lead to long-lived metastable states depending on initial conditions. We consider a model of interacting hardcore bosons with an additional kinetic constraint that was originally devised to capture glassy dynamics at high densities. As a main result, we demonstrate that the system is highly prone to localization in the presence of uncorrelated disorder. Adding disorder quickly triggers long-lived dynamics as evidenced in the time evolution of density autocorrelations. Moreover, the kinetic constraint favors localization also in the eigenstates, where a finite-size transition to a many-body localized phase occurs for much lower disorder strengths than for the same model without a kinetic constraint. Our work sheds light on the intricate 
interplay of kinetic constraints and localization and may provide  additional control over many-body localized phases in the time domain. 
\end{abstract}

\maketitle
\section{Introduction} 
In the theory of thermalization of closed quantum systems, two main pillars
have emerged that capture generic behavior. On the one hand, systems that obey the eigenstate thermalization hypothesis (ETH) are expected to thermalize under their own dynamics \cite{dAlessio2016,Gogolin2016,Deutsch2018,Mori2018}, i.e., information on initial conditions becomes inaccessible to local measurements and expectation values of local observables are identical to thermal expectation values, up to small finite-size corrections. On the other hand, many-body localized  (MBL) systems constitute robust examples of non-thermalization \cite{Nandkishore2015,Abanin2019}, with emergent local conserved quantities and persistent density inhomogeneities.

Two recent developments have triggered a refinement of this picture.
First, the stability of the MBL phase, even in a canonical system of a chain of interacting spinless fermions in presence of quenched disorder, has been challenged \cite{Devakul2015,DeRoeck2017,Doggen2018,Weiner2019,Panda2019,Suntajs2020,Suntajs2020a,Sierant2020,Kiefer-Emmanouilidis2020,Luitz2020,Kiefer-Emmanouilidis2021,Abanin2021,Sels2021,LeBlond2021,Crowley2022,Ghosh2022,Sierant2022,Sels2022,Evers2023,Sierant2023}. Notably, the critical disorder strength appears to be significantly larger than previously suggested and instead of a direct transition, the existence of a large prethermal MBL regime has been proposed \cite{Morningstar2022,Long2023}.
Second, quantum systems with various types of constrained dynamics have attracted significant attention, including systems with quantum scars \cite{Serbyn2021,Chandran2023}, Hilbert-space fragmentation (HSF) \cite{Khemani2020,Sala2020,Morningstar2020,Moudgalya2021}, lattice gauge theories \cite{Smith2017,Brenes2018,Karpov2021}, or kinetically constrained models (KCMs) (see, e.g., \cite{Lesanovsky2013,vanHorssen2015,Lan2018, Pancotti2020,Zadnik2023}).

Systems with constrained dynamics are interesting from several perspectives \cite{Garrahan2018,Serbyn2021,Moudgalya2022a,Chandran2023}.
In many cases, such systems eventually thermalize for the majority of initial
conditions, yet break ETH in the weak sense \cite{Serbyn2021,Moudgalya2022a}, however, even fully non-thermalizing situations have been suggested \cite{Brenes2018,Karpov2021}.
Notably, long transient dynamics and metastable states exist in these 
systems for at least some initial conditions, thus shifting the focus from 
eigenstate properties, as often emphasized in the ETH and MBL contexts, to the temporal domain.
The existence of quantum-scar states \cite{Turner2018} and HSF
in certain models has led to an improved view on local conservation laws responsible for these Hilbert-space structures \cite{Moudgalya2022}. Some of the models with HSF or kinetic constraints exhibit anomalously slow, subdiffusive transport 
\cite{Feldmeier2020,Iaconis2021,Singh2021,Burchards2022} or superdiffusive
transport \cite{Ljubotina2023}.
KCMs originate in two contexts, either in quantum systems with approximately hard-core short range interactions such as Rydberg atoms \cite{Lesanovsky2013} or in the classical theory of glassy dynamics \cite{Palmer1984,Fredrickson1984,Sollich1999,Ritort2003}. Transferring models
from the latter class into the quantum realm provides a rich playground to study the interplay of interactions, constraints, and disorder \cite{Lesanovsky2013,Causer2020}.
First nonequilibrium experiments with quantum simulators succeeded in 
demonstrating the presence of constrained dynamics \cite{Valado2016,Bernien2017,Letscher2017,Scherg2021,Zhang2022,Kohlert2023,Su2023}.

\begin{figure}
\centering
\includegraphics[width=0.8\linewidth]{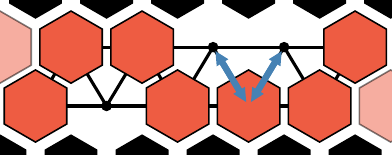}
\caption{Illustration of the kinetically constrained model of consideration in this work, introduced in \cite{Lan2018}. The black dots represent the sites on a  triangular (zig-zag) ladder. The red hexagons represent particles that can only hop to the nearest neighbor if the origin and target site share an empty neighboring site. The blue double-ended arrows denote the particle-hole interactions and the allowed hopping processes for this example configuration [see Eq.~\eqref{eq:ham}]. }
\label{fig:model}
\end{figure}

In our work, we are interested in the stability of localization in a KCM describing an interacting system of hard-core bosons or spin degrees of freedom on a triangular ladder (see Fig.~\ref{fig:model}).
In this system, introduced as a quantum version in Ref.~\cite{Lan2018}, particles can only move into an empty site if the origin and target site share an empty neighboring site (see the arrows in Fig.~\ref{fig:model}).
Consequently, single vacancies in a high-density background cannot move at all, unless they are absorbed by groups of at least two neighboring vacancies. This mechanism causes the existence of 
quantum-scar states that are completely isolated from the rest of the 
Hilbert space (those are the computational-basis product states with only isolated vacancies) and states with metastable dynamics as witnessed in density autocorrelations \cite{Lan2018} (see Fig.~\ref{fig:Ctwith}(b) for examples).
The constraint becomes effective in the presence of sufficiently
strong interactions between particles and {holes}. This interaction is chosen to realize a Rokhsar-Kivelson point in the many-body ground state of this model \cite{Lan2018}.

The kinetic constraints lead to the aforementioned metastable dynamics with slowly decaying autocorrelations computed from computational-basis product states, with an increasing fraction of such states as density grows. Consequently, the system has an inherent tendency towards localization. Since isolated holes can only propagate via high-order perturbative processes involving connected clusters of vacancies, a small amount of disorder will quickly affect these heavy objects.
In our work, we demonstrate two main results. First, even a small amount of disorder is sufficient to induce long-lived, nondecaying dynamics for  \textit{all} initial product states in the computational basis, at least on accessible finite system sizes. Second, the eigenstate transition into a possible many-body localized phase occurs at an order of magnitude smaller disorder strengths than for the model
\textit{without} the kinetic constraint, consistent with the information from time-dependent simulations. Our results are based on extensive exact-diagonalization simulations.

Our work complements  previous studies of constrained quantum-lattice models in the presence of disorder. For the case of KCMs, there is so far no uniform picture as the constraints can apparently favor or disfavor localization \cite{Chen2018,DeTomasi2019,Ostmann2019,Roy2020, Sierant2021,Herviou2021}, with our case providing an example for the former. It appears that the type of constraint, dimensionality and range of the 
interactions may matter.
A similar picture has been described for the quantum East random-energy model \cite{Roy2020}. There, however, localization is absent without the constraints
in the bulk of the spectrum \cite{Baldwin2016}. Moreover, the random-energy  model
is infinite-dimensional, different from our one-dimensional example.
With regard to the ongoing investigations about the stability of MBL, the combination of certain dynamical constraints and disorder may provide a path towards stable instances of MBL and possibly unexplored types of delocalization-localization transitions and crossovers.

The rest of the paper is organized as follows. In Sec.~\ref{sec:model},
we introduce the model while Sec.~\ref{sec:methods} provides a brief account of the numerical techniques utilized in our work.
In Sec.~\ref{sec:metastable}, we present our results for the decay of density autocorrelations
as a function of interaction strength and disorder strength.
Section \ref{sec:mbl} summarizes our results for the eigenstate delocalization-localization transition extracted from the occupation distance \cite{Hopjan2020,Hopjan2021}  and results for the eigenstate entanglement entropy. 
Our conclusions are presented in Sec.~\ref{sec:conclusions}.

\section{Model}
\label{sec:model}
Here we consider a triangular ladder with interacting particles subject to a kinetic  constraint introduced in \cite{Lan2018} and uncorrelated disorder. A schematic picture of the system is shown in Fig.~\ref{fig:model}. The system is governed by the Hamiltonian 
\begin{equation}
    \hat{H} = \hat{H}_{\text{KCM}} + \hat{H}_{\text{dis}}
\end{equation}
where
\begin{align}\label{eq:ham}
    \hat{H}_{\text{KCM}}  = & -J\sum_{\langle i,j \rangle} \hat  C_{i,j}  (\hat{b}_i^\dagger \hat{b}_{j} + \rm{h.c.}) \nonumber\\
    & + V\sum_{\langle i,j \rangle} \hat C_{i,j} \left\lbrack\hat{n}_i (1-\hat{n}_j) + \hat{n}_j (1-\hat{n}_i)\right\rbrack,
\end{align}
and
\begin{equation}
    \hat{H}_{\text{dis}} =  \sum_{i=1}^L \epsilon_i \hat{n}_i.
\end{equation}
Here, $\hat{b}_i^\dagger$ ($\hat{b}_i$) are bosonic creation (annihilation) operators subject to an onsite hardcore constraint and $\hat{n}_i$ are the number operators at a given site $i$, with $L$ the number of sites. The first term of $\hat{H}_{\text{KCM}}$ represents the  hopping with  amplitude $J$, and  $\hat C_{i,j} = 1- \prod_k \hat n_k$ defines the kinetic constraint where $k$ denotes all the common neighbor sites of $i$ and $j$. The particle-hole interaction is defined by the second term with interaction strength $V$, which is also subject to the constraint. $\hat{H}_{\text{dis}}$ stands for the disorder in the system where $\epsilon_i$ are  uniform random numbers drawn from a box distribution $ \epsilon_i \in [-W, W]$. $W$ is the strength of the disorder potential.

We will also consider a system \textit{without} the constraint, that is, a Hamiltonian $\hat H_{\text{un}}$ instead of $\hat H_{\text{KCM}}$ obtained from setting $\hat C_{i,j}=1$ in $\hat H_{\text{KCM}}$.
We define the filling as $\nu=N/L$, where $N$ is the particle number.

A central quantity in our analysis will be density autocorrelations $c(t)$, defined as
\begin{equation}
    c(t) = \frac{1}{L} \sum_{i=1}^L \frac{ \langle \psi(0)|\hat{n}_i(t)\hat{n}_i(0)|\psi(0)\rangle } {\nu(1-\nu)} - \frac{\nu}{1-\nu}\,,
\end{equation}
which we average over \textit{all} sites \cite{Lan2018}.
Here, $|\psi(0)\rangle$ denotes the initial state, which in our case
are always product states in the computational basis.

\section{Methods}
\label{sec:methods}
The Hamiltonian is implemented as the matrix representation in the basis of joint eigenvectors of the local density operators $\hat n_i$. In this form the constraint is just an on/off flag for each matrix element depending on the existence of neighboring vacant sites.

Time evolution and expectation values of energy eigenstates are obtained by exact diagonalization \cite{Manmana2005,Sandvik2010} where eigenvalue decomposition of the Hamiltonian is calculated using LAPACK \cite{lapack1999}.
The time average $\overline{c}(t)$ of the autocorrelation is obtained as a weighted average of values  at times $t_i J = 10^{\alpha i}$
with $0.05 \leq t_iJ\leq t$ and $\alpha\approx 0.04$ where the weight of a point at time $t_i$ is the length of the time interval $t_{i+1} - t_{i}$.
In principle, this leads to an overestimate of the long-time value,
which, however, does not affect our analysis since we are considering
the dynamics over many decades in time. Note that the plateaus in $c(t)$ without a time average are obscured by large temporal fluctuations \cite{Lan2018}.

Disorder averages of the time average of the autocorrelation are taken over samples of different disorder realizations and all initial configurations equivalent under symmetry transformation (mainly translation symmetry) in the clean model. The latter is just done to reduce the number of necessary disorder realizations.
The number of disorder samples is chosen such that the uncertainty of the average is smaller than  the linewidth in the respective plots.

\section{Long-lived dynamics in the presence of disorder}
\label{sec:metastable}

\subsection{Clean case}

\begin{figure}
\centering
 \includegraphics{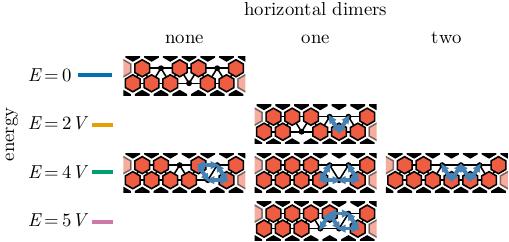}
\caption{Sketch of classes of initial states for $L=12$.
We distinguish the initial states by their interaction energy. For $E/V=4$, there are three different types of initial states. Those states with $V=0$  that have only isolated holes are exact quantum-scar states.}
\label{fig:initial}
\end{figure}

As mentioned above, $\hat{H}_{\text{KCM}}$ hosts metastable states in the  $V>J$ limit, where many initial states show a plateau in the density autocorrelation function $\overline{c}(t)$.
This behavior has been discussed in \cite{Lan2018}, which we here recapitulate to lay the ground for the discussion of the disorder case. 
We will focus on results for $L=12$ and a high filling of $\nu=3/4$. 

\begin{figure}
\centering
 \includegraphics{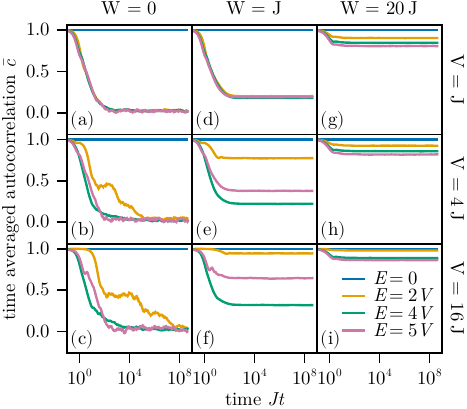}
\caption{Dynamics of the constrained model: $\overline{c}(t)$ for all initial product states  plotted for a system of $L=12$ 
sites for different interaction strength and disorder strength.
The results are averaged over the groups of  initial states according to the
classification from Fig.~\ref{fig:initial} and 20 disorder samples.}
\label{fig:Ctwith}
\end{figure}

\begin{figure}
\centering
 \includegraphics{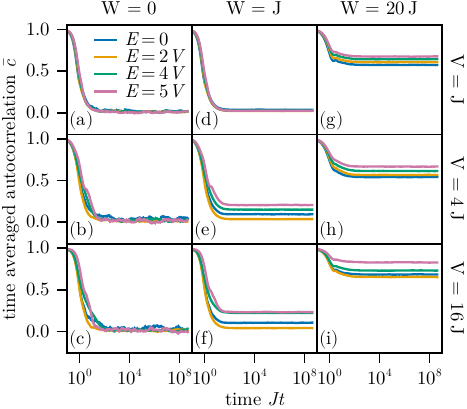}
\caption{Dynamics of the unconstrained model: $\bar c(t)$ for all  initial product states are plotted for a system of $L=12$ 
sites for different interaction strength and disorder strength without the kinetic constraint.
The results are averaged over the groups of initial states according to the
classification from Fig.~\ref{fig:initial} and 20 disorder samples.}
\label{fig:Ctwithout}
\end{figure}

Before going further, we want to illustrate the underlying physics of the existence of the plateaus in the relaxation dynamics, which also gives insight into the non-thermalizing behavior in the presence of small disorder strength. In Figs.~\ref{fig:Ctwith}(a)-(c) and \ref{fig:Ctwithout}(a)-(c), we
present the time evolution of the density autocorrelation function $\overline{c}(t)$ 
computed for the model with constraint and without constraint, respectively.
The results are computed for three interaction strengths, $V/J=1,4,16$.

As already shown in Ref.~\cite{Lan2018}, a sufficiently large ratio
of $V/J$ causes metastable dynamics, as clearly seen in Fig.~\ref{fig:Ctwith}(b). Specifically, $\overline{c}(t)$ develops a plateau for those initial states that involve one horizontal dimer and one isolated hole
(solid orange lines). Increasing $V/J$ leads to longer plateaus (note the logarithmic scale on the time axis). The emergence of this slow dynamics becomes particularly evident by comparison to the data for the model without a kinetic constraint, where such plateaus are absent.

Moreover, the KCM exhibits completely frozen states, namely those with only isolated holes, for which $c(t)=1$ for all times [see the solid blue lines in Figs.~\ref{fig:Ctwith}(a)-(c)]. These correspond to exact quantum-scar states with no hybridization with the rest of the spectrum and zero entanglement.
The density autocorrelations for all other initial states decay quickly on a time scale set by $(V/J)^2$ \cite{Lan2018}. 

\subsubsection{Perturbative estimate of time scales}
In order to guide the following discussion of the combined effect of disorder and interactions, we provide a discussion of the relevant time scales for the decay of the metastable states. These clearly depend on $V$ and the basic mechanism can be extracted from considering initial states with one horizontal dimer and one hole (see the state with $E=2V$ in Fig.~\ref{fig:initial}). 

For an isolated hole to be able to move, a horizontal dimer needs to first flip into a vertical dimer which involves an energy cost of $\Delta E = 2V$.  The vertical dimer can then propagate through the system and can absorb and reemit the hole, with eventually returning to the subspace with one horizontal dimer. 
There are several possible intermediate states that involve
three connected vacancies. The lowest-order process involves a state with a trimer and $E=4V$ (see Fig.~\ref{fig:initial}) and leads to a contribution in the  order of $J^4/V^3$. Going through an intermediate state with vacancies in a triangle leads to $J^5/V^4$. Note that the motion of a horizontal dimer itself in a background of
occupied sites goes with $J^3/V^2$.
These simple arguments are consistent with the dependence of the plateau width of the metastable states at $W=0$ [see
Fig.~\ref{fig:Ctwith}(b)]
on the interaction strength (not shown here).

In summary, the perturbation theory argument explains the dependence of the plateau length on $V$ and the sensitivity of different types of initial states to the constraint. Since we can therefore  view single holes as heavy objects with a small tunneling amplitude, the addition of disorder should lead to a rapid localization.

\begin{figure}
\centering
 \includegraphics{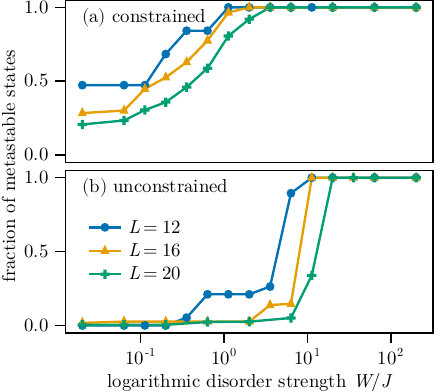}
\caption{ Fraction of metastable states as a function of $W$ for (a) constrained model and (b) unconstrained model for $L=12,16,20$ and $V=4J$. We consider the dynamics to be metastable when $\bar c(t_{\text{thresh}})>\epsilon$ with $\epsilon=0.15$  and $t_{\text{thresh}}=500/J$.  The results do not
quantitatively depend on the choice of $\epsilon$ and $t_{\text{thresh}}$ for reasonable choices of these parameters.}
\label{fig:meta}
\end{figure}

\subsection{Disorder and interactions}
We next discuss the effect of disorder on the time dependence of the autocorrelation functions.
These are shown in Figs.~\ref{fig:Ctwith}(d)-(i) and  Figs.~\ref{fig:Ctwithout}(d)-(i) for the constrained and unconstrained model, respectively. We show averages over those initial states that have the same
configurations according to Fig.~\ref{fig:initial}.
Remarkably, even  disorder strengths substantially smaller than the bare bandwidth $W\sim  J \lesssim 4J$ prevent the density autocorrelations from decaying over many decades for all initial states in the presence of constraints  [see Figs.~\ref{fig:Ctwith}(d)-(f)]. Increasing the disorder strength leads to a higher long-time saturation value of $\bar c(t)$. 
As we shall show later, the density autocorrelations do not decay at all for the system sizes
considered. Therefore, on finite system sizes, disorder leads
to nondecaying dynamics. We stress the difference to the definition of metastable dynamics used here which refers to long-lived correlations that eventually decay already on finite systems.

For the model without a kinetic constraint, there are noticeably less states that acquire 
plateaus in $\bar c(t)$ [see, e.g., Fig.~\ref{fig:Ctwithout}(e)], which also requires higher values of $W$. The difference between the two cases is best illustrated by plotting the fraction of metastable
initial states as a function of disorder strength shown in Fig.~\ref{fig:meta} for three system sizes
$L=12,16,20$ and both models at $V=4J$. 
We consider a state to be metastable when $c(t_{\text{thresh}})>\epsilon$
with $t_{\text{thresh}}=500/J$ and $\epsilon=0.15$.
In this analysis, we are sensitive to density autocorrelations
that remain large in the short-time regime, but at times larger than the generic decay time set by $(V/J)^2$ \cite{Lan2018}.
Clearly, the kinetic constraints lead to a short-time plateau of 
$\bar c(t)$ for an order of magnitude smaller values of $W$ than in the case without kinetic constraints. Increasing system size suppresses the metastable states in both cases, yet much more significantly so in the case without a constraint. 
%Based on the accessible system sizes, we cannot rule out that the 
%metastable states will decay for $L\to \infty$, yet our main point is the difference between constrained and unconstrained model.

Another noteworthy effect of the constraint is to
affect the plateau height of $\overline{c}(t)$  for the different groups of 
initial states as defined in Fig.~\ref{fig:initial}.
The comparison of, e.g., Figs.~\ref{fig:Ctwith}(b) and
Fig.~\ref{fig:Ctwithout}(b), shows that in the presence of the constraint, the states with one hole and one horizontal dimer are the most susceptible to disorder, while in the absence of the constraint, these states exhibit the lowest values and states with three vacancies in neighboring sites have the largest $\overline{c}(t)$ in the plateau. 
The sequence of plateau values for the unconstrained model $\hat H_{\text{un}}$ results from the interplay of available hopping processes versus the interactions that favors clusters of vacancies and is thus subject to details of the
values of $W$ and $V$.

\begin{figure}[t]
\centering
\includegraphics{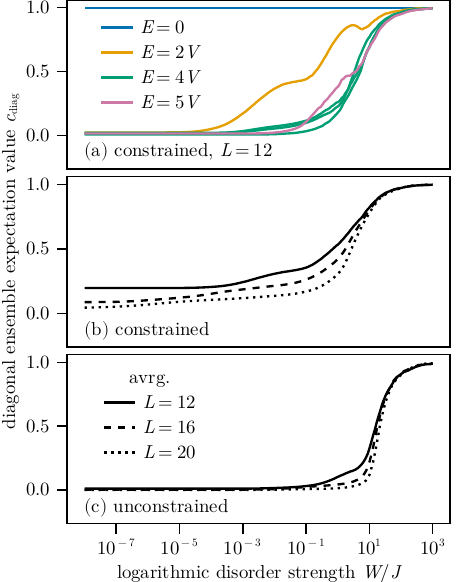}
\caption{Diagonal ensemble expectation value $c_{\text{diag}}$ of the density autocorrelator for (a)
$L=12$ for all groups of initial states. In (b) and (c) we show the infinite-temperature averages for $L=12,16,20$ for (b) the constrained model and (c) the unconstrained model.
}
\label{fig:diag_ensemble}
\end{figure}

One immediately wonders about the temporal extension of the plateaus in $\bar c(t)$ at $W>0$ \cite{Lan2018,Zadnik2023}
and whether they persist to infinite times as expected for
many-body-localization. 
While definite statements about the thermodynamic limit are difficult, we can 
 compute the diagonal ensemble expectation values \cite{Rigol2008} of the autocorrelator, $c_{\text{diag}}$. 
Figure~\ref{fig:diag_ensemble}(a) contains the data for $L=12$ and the different groups
of initial states.

The difference to the unconstrained model is the most obvious from
the comparison of Figs.~\ref{fig:diag_ensemble}(b) and \ref{fig:diag_ensemble}(c) that display the infinite-temperature average over all states for different system sizes.
 Using such data, we can demonstrate that there is no decay 
on finite systems (ad-hoc measured by $c(\infty) > 0.1$ for $L=20$ data) for $W/J\gtrsim 0.4 \cdot 10^{-4}$, as shown in Fig.~\ref{fig:diag_ensemble}(b) 
for the constrained model.  
For the unconstrained case [see Fig.~\ref{fig:diag_ensemble}(c)], the density autocorrelations acquire a nonzero long-time value for  orders of magnitude larger values of $W/J \gtrsim 7.74$. 
The dependence
of $c_{\text{diag}}$ on $W$ from Fig.~\ref{fig:diag_ensemble}(b) resembles  the one of the fraction of  states with metastable
shown in Fig.~\ref{fig:meta}(a) as it increases at around the same value of $W$. However, the diagonal ensemble misses information about metastable states on finite systems, which are captured in Fig.~\ref{fig:meta}(a).

The temporal and long-time behavior of $\overline{c}(t)$ is markedly different from
the stretched exponential decay of autocorrelations suggested for the prethermal-MBL regime \cite{Long2023}, which may thus be restricted to much smaller values of $W/J$. 
While here we focus on the emergence of non-decaying 
autocorrelations (averages and for individual initial computational
basis states), there is possibly another interesting regime at weaker disorder.  Given the different dynamics of initial states, some classes of those lead to non-decaying dynamics for smaller values of disorder than others. The remaining fast decaying states may be sufficient to ensure delocalization for all states eventually. Substantiating this scenario is left for 
future work.

%\subsection{Quantum Scars and Hilbert-space fragmentation}
%\fabian{tentative!}

\section{Localization-delocalization transition}
\label{sec:mbl}

So far, we have established that the kinetic constraints lead to 
non-decaying density auto-correlations for much smaller values of $W$
compared to the case without constraints.
We now complement this picture by studying the finite-size eigenstate
transition from a delocalized regime to a putative many-body localized regime.

To that end, we compute the occupation distance \cite{Hopjan2020,Hopjan2021} that is extracted from distributions of local densities $\langle \hat n_i \rangle $
sampled over sites, disorder realizations, and eigenstates. The distributions exhibit a 
bimodal structure in the localized regime, while they are normal-distributed around the average filling in the delocalized case \cite{Lim2016,Luitz2016}.

\begin{figure}
\centering
\includegraphics{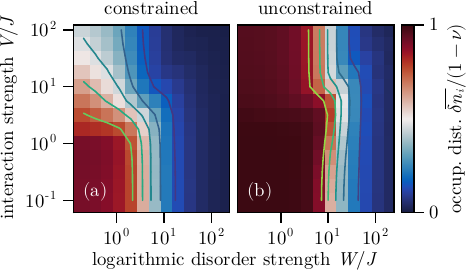}
\caption{Delocalization-localization state diagram in the $W-V$ plane extracted from the normalized  disorder-averaged occupation distance $\frac{\overline{\delta n_i}}{1-\nu}$, which is represented according to the color scale, for (a) the constrained  and (b) the unconstrained model, both for $L=16$. Here, $\nu = 3/4$ and the $\frac{\overline{\delta n_i}}{1-\nu}$ is averaged over $20$ disorder realizations, leading to a statistical variation of less than 0.03.}
\label{fig:pd}
\end{figure}

The occupation distance $\delta n_i$ is computed from each eigenstate expectation value of the onsite density $n_i = \langle \psi | \hat{n}_i |\psi \rangle$ as the distance from the closest integer
($|\psi\rangle$ denotes an eigenstate of the Hamiltonian). The definition reads
\begin{equation}
    \delta n_i = |n_i -[n_i]|.
\end{equation}
If the states are localized in nature, the average $\overline{\delta n_i}$ taken over disorder, sites and eigenstates goes close to zero and is practically system-size independent, while if the states are extended, $\overline{\delta n_i}$ approaches the filling $\nu$ for $\nu \leq 1/2$ or $1-\nu$ for $\nu > 1/2$ \cite{Hopjan2020,Hopjan2021}, where $\nu = N/L$.
The average of $\delta n_i$ over sites  and all  eigenstates for a given parameter set has been shown to capture delocalization-localization transitions \cite{Hopjan2021}, including the known critical behavior of transitions in non-interacting models.   

In Figs.~\ref{fig:pd}(a) and (b), we plot ${\overline{\delta n_i}}/(1-\nu)$, which is averaged over 20 disorder realizations, as a colour plot in the  $W-V$ parameter space for different system sizes  for $\nu = 3/4$. These state diagrams show that disorder affects the system differently for the constrained and unconstrained cases. In the former,  the transition sets in at   values of $W$ that are an order of magnitude smaller than in the latter case
[compare the lines in the figures, indicating equal values of ${\overline{\delta n_i}}/(1-\nu)$].

In the limit of small values of $V$, neither system becomes Anderson localized
for all values of $W$ since even in the absence of particle-hole interactions $V=0$, hardcore bosons are interacting particles. Large values of $V$ favor localization and therefore, the localized region wins over the delocalized one there.

\begin{figure}
\centering
\includegraphics{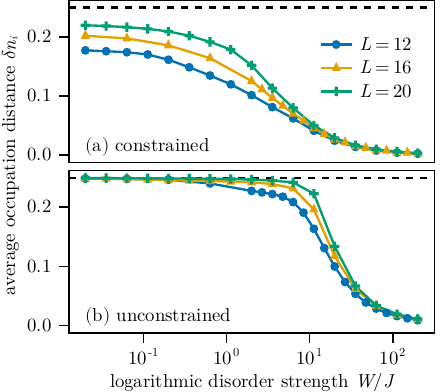}
\caption{System-size dependence of the disorder-averaged $\overline{\delta n_i}$ for (a) constrained model and (b) unconstrained model, both for $V=4J$ and $L=12,16,20$ sites averaged over (at least) $400, 30, 5$ disorder configurations, respectively.}
\label{fig:dn}
\end{figure}

The finite-size dependence of the occupation distance is illustrated in Fig.~\ref{fig:dn} for $V=4J$.
In the unconstrained model, the occupation distance already approaches its
asymptotic value of $\delta n_i =0.25$ in the range $W \lesssim 10J$, indicating
delocalization. In the constrained model, the data do not reach the limiting value anywhere yet, supporting the presence of  much larger finite-size effects and the presumably larger extension of the localized phase. 

\begin{figure}
\centering
\includegraphics{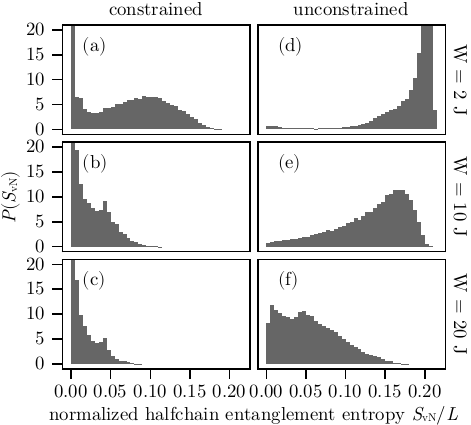}
\caption{Distribution of the half-chain entanglement entropy $S_{\text{vN}}$ computed in eigenstates and sampled
over 20 disorder realizations and eigenstates for $L=16$ and (a)-(c) constrained, (d)-(f)
unconstrained model for $V=4J$. We show results for $W/J=2,10,20$ and plot the data versus 
$S_{\text{vN}}/L$.}
\label{fig:svn}
\end{figure}

We have also studied the distribution of the half-chain entanglement entropy \cite{Kjaell2014}
computed in eigenstates.
The $S_\text{vN}$ is calculated from a bipartition of the system into $A$ and $B$ subsystems (here of equal length) and calculating the reduced density matrix of one of the subsystems ($\hat \rho_{A/B} = {\rm{Tr}}_{B/A}|\psi\rangle \langle\psi|$), as
\begin{equation}
    S_\text{vN} = -{\rm{Tr}}[\hat \rho_A{\rm{ln}}\hat \rho_A].
\end{equation}
Our results for the distribution $P(S_\text{vN})$ sampled over eigenstates and disorder realizations displayed in Fig.~\ref{fig:svn} further corroborate the
notion that the constrained model tends to localize much faster. Already at $W/J=2$, there is a broad distribution around a small mean value, with a larger additional peak at $S_\text{vN}=0$ stemming from the fully localized states. At $W/J=10$, 
$P(S_\text{vN})$ of the constrained model has the typical shape for a many-body localized system \cite{Lim2016}, with a maximum at $S_\text{vN}=0$, tails, and a local maximum at $S_\text{vN} \approx 0.7\approx \mbox{ln}\, 2$ (related to two-body resonances \cite{Lim2016}), while the unconstrained model still exhibits a 
broad distribution around much larger values of $S_\text{vN}$. Even at $W/J=20$, the unconstrained model does not show the sharp global maximum at small values of $S_\text{vN}$ yet.

Other quantities, such as  the gap ratio \cite{Oganesyan2007}, lead to the same picture
(results not shown here). The analysis of the gap ratio is, however, plagued by regions
with a vanishing density of states (that is, gaps in the many-body spectrum).

In conclusion, both the analysis of the time-dependence of autocorrelations
and of measures of a delocalization-localization transition yield
the same picture, namely, the kinetic constraints significantly favor localization. 
A scaling analysis of the stability of the localized phase is beyond the scope of our work and left for future work, similar to the case of the localized phase in the East-random energy model \cite{Roy2020}.

\section{Conclusions}
\label{sec:conclusions}

In our work, we provided numerical evidence that a certain type
of kinetic constraints in cooperation with interactions leads to an enhanced tendency towards localization in the presence of uncorrelated disorder. We established this result by primarily considering the time evolution of density autocorrelations corroborated by measures for an eigenstate localization transition such as occupation distance of density distributions and entanglement entropy. Our conclusion relies strongly 
on the direct comparison to particles that live on the same lattice topology yet are not subject to the kinetic constraints.
On finite systems, the crossover to localization occurs typically at an order of magnitude smaller values of disorder 
in the presence of the kinetic constraints, compared to when these are absent.

While our study is subject to the limitations of exact diagonalization and therefore, small system sizes, they still
suggest more stable localization in our KCM than in the absence of kinetic constraints. Putting this onto more theoretical grounds and on a broader data basis in terms of examples of KCMs is left for future work.

Based on our results, it seems likely that other mechanisms will also render many-body systems more
susceptible to disorder, at least in the sense of long-lived correlations in the  time domain.
These include flat-band systems \cite{Danieli2020,Kuno2020,Daumann2023}, systems with frustration \cite{McClarty2020,Bahovadinov2022}, and systems with emergent particle excitations with a narrow bandwidth such as heavy fermions or polarons 
in electron-phonon systems \cite{Jansen2019,Schoenle2021}.

We acknowledge useful discussions with K. Hazzard, I. Lesanovsky, D. Luitz, and L. Vidmar. 
This work was funded by the Deutsche Forschungsgemeinschaft (DFG, German Research Foundation) –  499180199, 436382789, 493420525 via  FOR 5522 and large-equipment grants (GOEGrid cluster).
This research was supported in part by the National Science Foundation under Grant No. NSF PHY-1748958. 

The data shown in the figures will be made available as ancilla files
on the arxiv. 

%\bibliography{references}
%apsrev4-2.bst 2019-01-14 (MD) hand-edited version of apsrev4-1.bst
%Control: key (0)
%Control: author (8) initials jnrlst
%Control: editor formatted (1) identically to author
%Control: production of article title (0) allowed
%Control: page (0) single
%Control: year (1) truncated
%Control: production of eprint (0) enabled
%

\end{document}